# An Entropy Analysis based Intrusion Detection System for Controller Area Network in Vehicles

Qian Wang*, Zhaojun Lu †, and Gang Qu *
* Department of Electrical and Computer Engineering, University of Maryland, College Park, United States
† School of Optical and Electronic Information, Huazhong University of Science and Technology, Wuhan, China
E-mail: {qwang126, gangqu}@umd.edu {D201377521 }@hust.edu.cn

*Abstract*—Dozens of Electronic Control Units (ECUs) can be found on modern vehicles for safety and driving assistance. These ECUs also introduce new security vulnerabilities as recent attacks have been reported by plugging the in-vehicle system or through wireless access. In this paper, we focus on the security of the Controller Area Network (CAN), which is a standard for communication among ECUs. CAN bus by design does not have sufficient security features to protect it from insider or outsider attacks. Intrusion detection system (IDS) is one of the most effective ways to enhance vehicle security on the insecure CAN bus protocol. We propose a new IDS based on the entropy of the identifier bits in CAN messages. The key observation is that all the known CAN message injection attacks need to alter the CAN ID bits and analyzing the entropy of such bits can be an effective way to detect those attacks. We collected real CAN messages from a vehicle (2016 Ford Fusion) and performed simulated message injection attacks. The experimental results showed that our entropy based IDS can successfully detect all the injection attacks without disrupting the communication on CAN.

## I. INTRODUCTION

In the past decade, more and more advanced functions and features have been added to modern vehicles which make them not only safer but also connected, smarter, and more intelligent. To support these growing functions and features, a large amount of software and hardware electronic equipment is integrated into the car system. For example, with the recent aggressive push of autonomous vehicles, many Electronic Control Units (ECUs) are introduced for data processing and communications between inside the vehicle network and the Vehicle Ad-hoc Network (VANET) [1]. Moreover, the applications of the vehicle network make the vehicle as a connected device to the Internet of Things (IoT), not just a traditional physical part. As vehicles adopt more connectivity functions to the external network, security threats on electronic controllers of vehicles are highly rising. With safety remains as the most important concern for automakers, security is now gaining lots of attention because a compromised ECU or faked communication messages can cause fatal failures just like mechanical problems. It has been shown that attacks can be successfully launched on cars to cause severe impacts on the normal driving behaviors [2], [3]. When the in-vehicle communication network was designed in the late 1980s, there were only about 1% electronic equipment in the vehicle system. Thus the original design requirements of such network at that time were merely to connect the limited in-vehicle ECUs. A simple message broadcasting based communication protocol, known as the Controller Area Network (CAN), was used for this purpose without any consideration of security. Not surprisingly, due to its lack of security concerns and protection, CAN's vulnerabilities are exposed to the attackers and become a major target for a variety of threat models. The successful attacks demonstrate that without the intrusion detection or authentication methods, it becomes very challenging for the CAN-based communication network to defend emerging attacks such as message replays, injections, and modification [4], [5]. Nevertheless, the important role that CAN plays in the communication of vehicular systems has made it an irreplaceable component and thus the research community has continued its efforts in seeking secure and efficient protection method for the CAN network.

Generally speaking, there are two kinds of countermeasures to defend against attacks on CAN bus: message authentication and intrusion detection. Message authentication code (MAC) has been successfully used in other security applications, but it is highly restricted by the nature of CAN that was not designed for security. For example, the data length of the regular CAN message is only 8 bytes which limits the space for appending a message authentication code. Researchers have demonstrated that several MAC-based security schemes could provide protections to the CAN bus [6], [7]. However, these schemes have limited flexibility in adapting to new applications and incur high cost in the MAC deployment. Furthermore, inline security protection methods like message authentication often complicate the communication and result in significantly increased overhead in performance such as the response delay. The real-time requirements on the vehicle system for time-critical applications such as braking system would pose a high demand on the processing time to compute, append, and verify the MACs.

Because of the limitation of MAC, Intrusion Detection System (IDS) has become a better choice. The IDS will work as monitoring the contents and the periodicity of in-vehicle CAN message and verify whether there are any significant changes. In [8], the authors proposed an entropy-based anomaly detection method for in-vehicle networks. They consider the scenario that an attacker injects messages with all 0's (which dominant other messages in CAN bus arbitration) into CAN bus. Computing message entropy can effectively detect such attacks. However, they fail to consider other complex attack scenarios. For example, the attacker might inject some purposely generated non-all-0 high priority messages to obtain control of the CAN bus. Among the other anomaly-based IDS methods, the clock-based IDS uses the clock skew to fingerprint the ECUs [9]. However, this method would need off-line computations to extract the fingerprint for each ECU which is difficult to meet the requirement for real-time detection.

In this paper, we focus on building a more effective method to detect a variety of injection attacks on CAN bus. The main idea is based on the observation that in the CAN message arbitration protocol, 0 dominates 1, which means that message with bit 0 will win the CAN bus over message with bit 1. Therefore, any injected messages, seeking to block other messages from accessing the CAN bus, would change the entropy at the bit level. By studying which bit's entropy changes, we develop a fast and efficient algorithm to search for the malicious ID as well. Further, the malicious messages containing those IDs would be discarded or blocked. More specifically, this paper makes the following contributions:

1) We develop a novel intrusion detection method based on the entropy changes of the CAN ID, which can be used to alert the CAN bus system. This method is designed to cover all known CAN message injection attacks.

2) We implement the proposed IDS and validate its effectiveness on car data collected from a Ford Fusion internal CAN network. And we simulate the attacks on the CAN bus prototype by using the Arduino UNO microcontroller and CAN bus Shield.

3) Unlike MAC based protection method that needs to modify the CAN messages, our entropy based IDS is a passive mechanism and it is independent of whether or which cryptographic algorithm is used. In other words, our approach will not modify and is applicable to any implementation of CAN protocols, including those modified with security policies such as MAC.

The rest of the paper is organized as follows. In section II, we give the necessary backgrounds on the CAN message format and arbitration protocol. In Section III, we propose two adversary models based on the attacking capability of the attacker. We elaborate our entropy based IDS in Section IV and validate it in Section V before concluding in Section VI.

## II. BACKGROUND: CAN BUS PRELIMINARIES

### A. CAN Frame

CAN is the most widely deployed in-vehicle communication protocol, which interconnects ECUs through a multi-master, message broadcast bus system. The involved ECUs communicate with each other on the CAN bus without the host computer to do the arbitration. The CAN frame contains fields such as Identifier (ID), Data Length Code (DLC), Data field, CRC and other control bits as shown in Fig.1. Since CAN is a message-oriented protocol, there are no transmitter or receiver addresses in the data frame. Nevertheless, a CAN frame contains a unique ID which represents the priority of the message and the function of the message. But without the source and target information, CAN messages are easily forged and difficult to be traced.

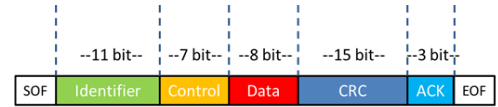

Fig. 1. The format of the CAN data frame

### B. CAN Bus Arbitration

CAN data transmission uses a bitwise arbitration method to decide which message could be sent on the bus when multiple ECUs request the bus simultaneously. The CAN specification defines the dominant bits (logic 0) and recessive bits (logic 1) based on the voltage level. If one node transmits a dominant bit and another node transmits a recessive bit simultaneously, the node with a dominant bit would win the arbitration. The arbitration is done in the ID field of the frame. That is to say, those frames with higher priority in ID would be transmitted before the others. And the frames with lower priority which fail to transmit last time would automatically attempt to re-transmit six clocks after the end of the last message. This makes CAN bus system very suitable as a real-time prioritized communication system.

## III. ADVERSARY MODELS AND ATTACK SCENARIOS

### A. Strong and Weak Models

Depending on the hardware, software and attack surfaces, ECUs would achieve different degrees of attack capabilities. Here, we will consider two different types of adversary models: strong and weak.

For the weak adversary model, the comprised ECU can only send out messages with the specific ID assigned to him. That is to say, the weak attacker cannot arbitrary select CAN IDs for the malicious messages. This mechanism relies on the transmitter filter installed outside of the ECU to stop the malicious messages without valid ID sending on the CAN bus. To hamper the communications on the CAN bus, a weak attacker can only send messages with limited IDs which could pass the filter. However, the attacker would still fulfill his malicious role by sending out junk or malicious messages with specific IDs, if those IDs will win the bus arbitration.

In contrast, the strong attacker could send out messages with any IDs. In this case, the strong attacker hampers the communications on CAN bus for two purposes. First, he would send malicious messages with higher priority IDs (leading zeros are in the front, e.g. 0000XXX) to override any periodic messages sent by a legitimate safety-critical ECU. Those malicious messages might be targeted to victim ECUs. As a result, the receiver ECU would get distracted, inoperable or execute errors. Second, by sending messages with higher priority IDs, the attacker ECU could occupy the CAN bus for a period of time and stop the legal ECUs sending their regular messages, which indeed hinders the regular operations of the vehicle. In other words, the adversary could maliciously stress the bus and compromise the usability of the other users just like flooding attacks. Normal flooding attacks can be easily detected by the network filters on the bus gateway or the dominant check mechanism of the CAN transceiver. However, the strong attack model discussed in this paper is more powerful than the normal flooding Denial of Service (DoS) attack. By holding the bus for a time, the attacker could simultaneously put his malicious messages on the bus which will cause hazardous effects to the vehicles. Thus, it is necessary to find out an approach to detect the attacks and stop it in the future transmission. The practicability of such an adversary model has already been proved and demonstrated in [9, 10].

*B. Specific Attack Scenarios*

In the following, we will describe four specific attack scenarios to test the effects of our entropy based IDS. All presented attack scenarios are derived from those well-known attacks on the in-vehicle network which have been demonstrated in previous research publications [9, 10].

*1) Strong Adversary Model: Flooding Message Injection*

In this scenario, the attacker has strong control of a compromised ECU and aims to break the availability of the CAN bus by sending out massive messages on the bus. This attack is usually deployed by injecting CAN messages containing the most dominant identifier, i.e. 0x00. However, the CAN transceivers have the detection mechanism for zero overloads on CAN bus. Specifically, if the transceiver sees the bus is occupied by 0 (high voltage), it will automatically shut down the transmission by pulling the high voltage to low. An efficient strategy for the attacker to avoid the detection mechanism is sending messages with changeable IDs of high priority. This kind of flooding attack is difficult to be detected by the transceivers. Furthermore, because no specific receivers are targeted for those malicious messages, the abnormal traffic would not be reported by the victim receivers. For this attack, the attacker can manipulate the effects of the system and the exposure of the detection system. The attacker can choose to send flooding messages with changeable IDs to overwhelm the regular communications. However, it will cause significant changes on the CAN bus, thus makes the attack easily to be detected. Consequently, the compromised ECU would be spotted because of too many junk messages, it will be shut down and cannot mount attacks on the CAN bus anymore.

*2) Strong Adversary Model: Message Injection with Single ID*

In contrast to the flooding scenario discussed above, the attacker could narrow down the number of IDs he uses to conceal himself from the detection schemes. Let us suppose the simplest one, the attacker would inject malicious messages with just a single CAN ID for two purposes: first, he wants to win the bus arbitration from the other low priority IDs thus stopping the transmission of useful messages on the bus; second, the malicious attacker could alter the contents of messages and send wrong information out with the fixed ID which would disturb the functions of the receiver ECUs.

The strategy for attacker may change based on his purpose. If he only focuses on disturbing the CAN bus, he can arbitrarily select the injected ID. Otherwise, if he wants to achieve both, he needs to pick up the injected ID from the legal ID set of the vehicle. The ID of CAN bus is assigned by the vehicle manufacturer and only around 10% to 20% of total number of IDs is used. For example, for the 2016 Ford Fusion we tested, only 223 IDs are used, which takes 10.88 % of the whole ID set.

*3) Strong Adversary Model: Messages Injection with Multiple IDs*

The same attack set-up as the scenario 2, but this time we assume that there are multiple attackers with different injected IDs, or the attacker can inject messages with multiple IDs. That is to say, the balance of entropy would be disturbed by two or more CAN IDs. This makes detection easier but inferring the faked IDs more difficult because multiple sources would influence the balance of the bus simultaneously.

*4) Weak Adversary Model: Messages Injection with Fixed ID*

In this scenario, we assume that the attacker can get the control of the ECU, but fail to break the transmitter filter. In this situation, the attacker can only transmit the malicious messages with fixed IDs assigned to the compromised ECU. Even though the control power is restricted in the weak model, the attacker can still break the availability of the CAN bus by sending out malicious messages. That happens when the ID assigned to that ECU are dominant the original messages transmitted at this time. By manipulating the frequency, the attacker can achieve to gain the bus control using a limited number of CAN IDs.

## IV. FRAMEWORK OF ENTROPY-BASED INTRUSION DETECTION SYSTEM

*A. ID and Binary Entropy Definition*

The identifier (11 bits) in CAN frame represents the message's priority as well as its functions, but without any transmitter or receiver information. One ECU could have multiple assigned IDs for different functionalities and several ECUs might be allowed to transmit the same ID, but not at the same time. More specifically, an ID is not unique for one ECU and could be consumed by different ECUs as well. Indeed, every packet in a vehicular CAN network and its possible data content are specified. The permitted value range is defined as well as the length of every signal and the packet function. The basic frame format has an 11-bit ID, whereas an extended format has a 29-bit ID. We will use the basic CAN ID format (11-bit) because it is much more prevalent, however, the detection method proposed in this paper is not dependent on the type of format, and it could also be applied to the extended format. In this paper, we use the probability of '1' occurrence at each bit to represent the distribution of that bit. As defined in the following,

**Definition:** $p_i$ is the number of messages where bit $i$ is 1 over the total number of messages. Notably, the range of $p_i$ is from [0,1]. The binary entropy function $H(p)$ is defined as the entropy of a Bernoulli process with probability $p$. If $\Pr(X=1) = p$, then $\Pr(X=0) = 1-p$ and the entropy of $X$ in Shannon is given by

$$H(X) = H_b(p) = -p\log_2 p - (1-p)\log_2(1-p).$$

## B. Entropy Detection Scheme

The first step is collecting data during normal driving to generate the golden template for detection. Because most of the messages on CAN bus are periodic, the distribution of entropy should be steady during normal driving. Data collected from a real Ford Fusion CAN bus demonstrate that assumption. The raw data from CAN are collected from different driving situations, e.g. turning the audio on, turning the light on, and driving with cruise control and so on. We observe that the entropy on each bit only changes slightly in these different testing scenarios. The variation falls in the range from $1e^{-8}$ to $9e^{-8}$ which is quite small. Therefore, the assumption as the entropy distribution under normal driving is steady is valid. Significant changes from the golden template will be treated as under attack.

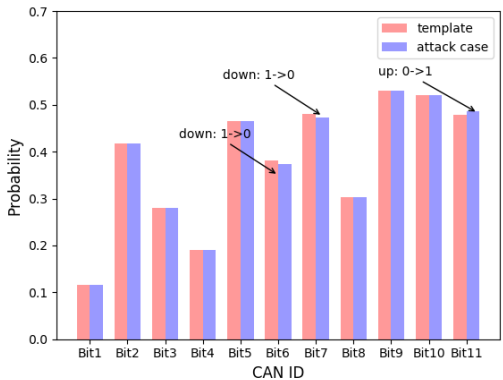

Fig. 2. Golden template and a case study example of an attack

Let $\hat{\mathbf{H}}=\{H_1\ H_2\ H_3\ldots H_{11}\}$ be the binary entropy vector measured from the real CAN data. We get the golden template $\mathbf{H_{temp}}$ by averaging of 35 measurements from diverse driving behaviors. For each bit in $\mathbf{H_{temp}}$, we calculate the range as $\max(H_i)-\min(H_i)$. And the threshold for detection can be set as a positive multiple of the range, expressed as $Th=\kappa\ (\max(H_i)-\min(H_i))$, where $\kappa$ is the coefficient which determines the margin of the threshold. The value of $\kappa$ is chosen from $\kappa=[3,10]$ empirically. For the experiment in this paper, we choose $\kappa=5$. The golden template as an 11-bit vector is shown in Fig.2. In the detection procedure, we compare the binary entropy to the template bit by bit. If the bit change is above the threshold, we will treat the CAN bus is under intrusion attack, and the system will send an alert signal. Fig.2 also illustrates one example of entropy changes under the injection attack. It is evident to find the significant changes occurred at some bits, e.g., Bit 6, Bit 7 and Bit 11.

## V. EVALUATION AND RESULTS

In this section, we first discuss and define the metrics used to evaluate the detection system. And then we present our experimental results of different attack scenarios, which demonstrate that our entropy detection system is effective on all the types of attacks.

### A. Test Set-up

For the information-theoretic measures introduced above, access to the vehicle network is required. In our set-up, we use the Vehicle Spy 3 Professional software tool to connect a 2016 Ford Fusion internal CAN network through the OBD-II port. And the baud rate is set as the standard for the CAN network: 125 kbit/s for the middle-speed CAN and 500 kbit/s for the high-speed CAN. We conduct our tests and experiments on the data acquired from the middle-speed CAN bus in this paper, however, our detection method would also work for high-speed CAN bus.

### B. Injection Rate and Detection Rate

First, we will introduce the metrics used in this paper. We define the injection rate $Ir$ as the proportion of successfully injected messages on the bus over the total number of messages the malicious ECU sends to compete for the bus arbitration. The injection rate is relevant to the capability of specific ID to win the CAN bus. As the CAN bus's arbitration relies on the ID priority, different IDs would win the bus arbitration with different probabilities. The number of times the attacker win the bus over the total times he tries in a fixed period should be a good metric to evaluate the injection power of the ID, which is the injection rate. Moreover, if the attacker wants to accomplish the attack, he needs to select the proper ID to win the bus arbitration first, which in turn results in increasing the number of injected messages overall.

We calculate the injection rate defined above for some selected IDs from the CAN log data and the results are shown in Fig.3. From the figure, we could find out that the injection rate is high when the ID values are small and it will drop down with the value increasing. That is because CAN bus does arbitration on bit level and bit 0 always dominants bit 1. The injection rate is significant as it represents the power of the attacker and also highly relates to the detection rate which we will explain in the following.

The detection rate $Dr$ is defined as the proportion of successfully detected injected messages over the total number of injected. The detection rate is the proportion of malicious messages that are correctly identified. Obviously, the detection rate is highly related to the entropy changes on the CAN bus. And the entropy change is influenced by the number of injected messages. Further, the number of injected messages depends on three factors: the injection rate $Ir$ of a specific ID, the injection frequency $f$ and the period of time $T_0$. Thus, the total number of successfully injected messages ($N_m$) can be expressed as a product shown in the formula

$$N_m = Ir \times f \times T_0$$

If the malicious attacker attempts to increase the success rate of injection during a fixed period of time, he should either choose the ID with a higher injection rate or increase the injection frequency. To demonstrate the relationship between injection rate and detection rate, we test 15 selected IDs under the same injection frequency. As the results are shown in Fig.3, the detection rate (blue line) decreases along with the injection rate. That is because the total number of injected messages on the CAN bus would decrease which makes the entropy changes imperceptible.

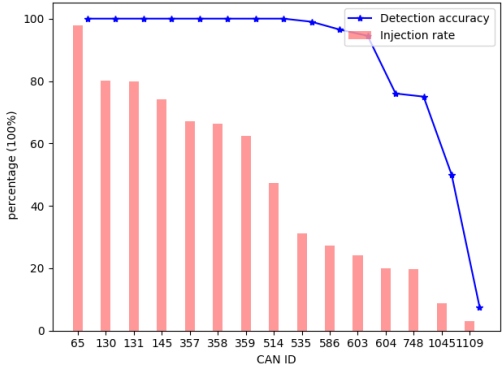

Fig. 3. Injection and detection rate for different CAN ID

## C. Inferring Malicious IDs

Generally saying, our Intrusion Detection System has two main tasks: 1) Detect the changes and demonstrate the evidence of the attack; 2) Figure out the injected message IDs and alert to the system for those malicious IDs. Here, we will discuss how to infer the malicious IDs based on the changes of the entropy observed on CAN bus.

For example, if the bit entropy changes in the negative direction (from 1 to 0), we can infer that the corresponding bit of the injected ID will be probably 0. Then, we can select those IDs in the ID pool with that bit is 0. When conducting this selection procedure bit by bit, the candidates' space will narrow down by combining the observing changes. However, this inference is non-deterministic as we have no clues on the other unchanged bit, which could be either 0 or 1. As a result, we propose the rank selection method to evaluate the accuracy in searching the malicious IDs. The procedure works as follows, we select the first $n$ rank candidates which obey the constraints derived from the entropy changes. If the malicious ID is in the candidate set, we will mark this detection as a hit. We use the hit rate to evaluate the accuracy of inferring. Due to the arbitration rule in CAN bus, the ID would be more powerful in injection if the most significant bits are 0. As a result, we sort the IDs in numeral ascending order, which means the preceding IDs in the sorted list would have a higher probability to be the malicious ID. The results of inferring accuracy for different attack scenarios are shown in Table I. And for the inference in our experiment, we set rank=10, which indicates each time we select the first 10 IDs as the candidates.

## D. Results Analysis of Different Attack Scenarios

We evaluate the feasibility of the proposed Intrusion Detection System on the four attack scenarios discussed above. First, we evaluate the detection rate of different scenarios to verify the accuracy and efficiency, and then we discuss on the inferring rate for the malicious ID under different attack scenarios.

The four test scenarios are listed as the flooding injection (FI), single message injection (SI), multiple messages injection(MI), and restricted ID message injection in weak mode (WI) with different injection frequencies(i.e., 100Hz, 50Hz, 20Hz and 10Hz). The detection rates for different scenarios are summarized in Table I.

In the flooding attack scenario, the attacker sends out many faked messages on the bus. It would cause dramatic entropy changes and this attack could be detected quickly. However, as the attacker arbitrary chooses the injected IDs without clear preference, it is infeasible to infer the injected IDs based on the almost randomly changed entropy. But if the attacker keeps sending flooding messages with different IDs, it will be easily detected by the filter in the gateway. Besides, it will leak some side-channel information to the internal detection sensor of this system as it is mounted on the bus for such a long period of time. For example, the power consumption and the requests for communication will be distinct for this compromised ECU.

For the single message injection, the attacker injected messages with one specific ID to influence the function of the target ECU. First of all, this injection would cause entropy changes on the CAN bus and our intrusion system would have on average 91% rate to detect this attack. Note this result is the average on every test CAN IDs including those with lower injection rate. For those CAN IDs with a higher injection rate, the detection rate could be as high as 100%. In this scenario, it is easier to infer the malicious ID based on the entropy changes, so the inferring accuracy is as high as 97.2%.

Then, we test and evaluate for multiple ID injections. In this case, messages with multiple CAN IDs are injected into the CAN bus causing complex changes than a single message injection. The basic approach to detect the multiple injection messages starts with inferring the ID based on a combination of the entropy changes. To achieve the purpose, we need to analyze not only the change direction but also the changing rate of each bit. We modify the selection algorithm to derive new constraints. Compared with the results for the single injection, the average detection rate is higher because multiple IDs are used with the increasing of the injection rate. However, the accuracy of inferring malicious IDs decreases when having multiple injected IDs. That is reasonable as the disturbance of multiple sources makes the algorithm difficult to select the exact combination of malicious IDs. Moreover, we test the multiple injection attacks with 2, 3 and 4 injected IDs respectively and the injection rate keeps going up as we enlarge the number of IDs while the inference accuracy goes down because of the disturbance of multiple IDs. As we mentioned above, the multiple ID injections are restricted by the filter and in real applications, it is challengeable for the attacker to have a fully-compromised ECU on the CAN bus. With 4 and more injection IDs, the compromised ECU would be easily figured out by the gateway filter.

We also complete the same test on the weak adversary model, which the attacker is restricted by the sender filter. The attacker can only send the messages with the ID assigned to him. The detection result is same as the single message injection, however, the inferring accuracy is a bit lower than the single ID injection, because the attacker is not restricted to inject single ID on the bus, and he can manipulatively change the entropy by using multiple IDs he could legally send.

From the results discussed above, we conclude that the inferring accuracies for several scenarios are above 90% as well. With our golden entropy model, we are able to successfully detect almost all the attacks (the 2nd column in Table I) and locate the malicious message ID except the flooding attack (the 3rd column in Table 1).

TABLE I. EVALUATION RESULTS FOR DIFFERENT ATTACKS

| Attack scenarios | Detection rate | Inferring accuracy |
|---|---|---|
| Flood | 100% | -- |
| Single Injection | 91% | 97.2% |
| Multiple_Injection_2 | 97% | 91.8% |
| Multiple_Injection_3 | 97.2% | 88.5% |
| Multiple_Injection_4 | 99.97% | 69.7% |
| Weak Injection | 93% | 96.6% |

### E. Advantages of the Proposed Approach

Lots of IDS have been discussed so far to protect the CAN bus communications [8,11]. We will compare our method with two representatives IDSs in this section.

In [8], the authors proposed an entropy-based approach for intrusion detection. However, their approach studies on the entropy changes of the whole message and treat the 11-bit ID as an inseparable vector. In our work, we split the 11-bit ID vector and analyze the entropy change bit by bit. Our method has several advantages over the approach in [8]. First, we could take use of the changes on the bit level to infer the malicious ID which is not implemented in their approach. Because the most significant bit has higher priority and much more influence on the arbitration, we will first focus on the higher priority bit. Besides, our bit-level entropy calculation would achieve a relative saving in the computing the entropy (from hundreds of elements down to 11) which is a benefit for the real react system for vehicles. The IDS for network security in vehicles always requires a light-weight detection algorithm because of the limitation of the computing power of electronic devices in cars. In [8], they need to count and store the occurrence of every ID on the CAN-bus for a relatively long period of time to determine the injection. Our system is designed to react quickly in a time period of as short as 1s due to the saving in the calculation. Besides, each ID in the set would require a memory space, which is not a negligible cost as the growing ID numbers in the network, especially for the constrict resources in the embedded system. However, in our bit-slice method, we just need 11 memory spaces to store the counting results for each bit, no matter how many ID messages are on the bus.

In [11], the authors built an Intrusion Detection System based on the analysis of time intervals of CAN messages. Compared to their method, our entropy-based method also supers as the cost-effective storage. In the time interval based detection method, they need to monitor the changes of the period for all the signals transmitting on the bus. As we stated before, each ID needs a specific storage space for the past and current period, thus introducing linear storage consumption. In addition, their method only records the period changes of specific ID, it cannot figure out such an attack scenario when the attacker uses unseen ID.

detection rates are all above 90% for all the scenarios. And the

## VI. CONCLUSION

We have developed a bit entropy based intrusion detection system which is a lightweight and nonintrusive security framework that can protect CAN bus without modifying its protocols and implementation. Our system is capable of restricting attackers from injecting a large number of malicious messages with higher priority IDs, as well as finding out the malicious IDs. These claims are supported by data collected from a running vehicle and simulated attacks.

**Acknowledgement.** Qian Wang and Gang Qu are supported in part by AFOSR MURI under award number FA9550-14-1-0351. Zhaojun Lu performed the work while he is visiting the University of Maryland, College Park.